%% file: paper_accepted.tex
\documentstyle{mn}

\newif\ifAMStwofonts

\input psfig2
\def\hh{$^h$}
\def\mm{$^m$}
\def\ss{$^s$}
\def\deg{\degr}
\def\asec{\arcsec}
\def\amin{\arcmin}
\def\etal{rm et al.}
\def\Msol{M$_{\sun}$}
\def\Msol{M$_{\sun}$}
\def\Mjup{M$_{\mathrm{JUP}}$}
\def\Rjup{R$_{\mathrm{JUP}}$}
\def\Msini{M\,$\sin i$}
\def\Lsol{L$_{\odot}$}
\def\kms{km\,s$^{-1}$}
\def\ms{m\,s$^{-1}$}
\def\sqde{$\Box$\deg}
\def\II{{\sc{II}}}
\def\rhk{R$^{\prime}_{\mathrm HK}$}
\def\rhkMW{R$^{\prime}_{\mathrm HK,MW}$}
\def\rhkCTIO{R$^{\prime}_{\mathrm HK,CTIO}$}
\def\rhkAAT{R$^{\prime}_{\mathrm HK,AAT}$}

\title[Echelle spectroscopy of Ca {\sc II} HK activity]
      {Echelle spectroscopy of Ca {\normalsize II} HK activity in Southern
       Hemisphere planet search targets\thanks{
       Based on observations obtained at the
       Anglo--Australian Telescope, Siding Spring, Australia.}}
\author[C. G. Tinney et al.]
       {C. G. Tinney,$^1$, Chris McCarthy,$^2$ Hugh R.A. Jones,$^3$
       R. Paul Butler,$^2$ \newauthor
       Brad D. Carter,$^4$
       Geoffrey W. Marcy,$^{5,6}$ Alan J. Penny,$^7$
        \\
 $^1$Anglo-Australian Observatory, PO Box 296, Epping. N.S.W. 1710. 
     Australia. {\tt (cgt@aaoepp.aao.gov.au)} \\
 $^2$Carnegie Institution of Washington,Department of Terrestrial Magnetism,
     5241 Broad Branch Rd NW, Washington, DC 20015-1305. USA.\\
 $^3$Astrophysics Research Institute, Liverpool John Moores University, 
       Twelve Quays House, Egerton Wharf, Birkenhead CH41 1LD, UK\\
$^4$Faculty of Sciences, University of Southern Queensland, Toowoomba, 4350. Australia.\\
$^5$Department of Astronomy, University of California, Berkeley, CA, 94720. USA.\\
$^6$Department of Physics and Astronomy, San Francisco State University, 
    San Francisco, CA 94132. USA.\\
$^7$Rutherford Appleton Laboratory, Chilton, Didcot, Oxon OX11 0QX, U.K.   }
\date{Accepted ---.
      Received January 10, 2002;
      in original form October 24, 2001}

\pagerange{\pageref{firstpage}--\pageref{lastpage}}
\pubyear{2002}

\begin{document}

\maketitle

\label{firstpage}

\begin{abstract}

We present the results of ultraviolet echelle spectroscopy of a 
sample of fifty nine F,G,K and M stars from the Anglo-Australian Planet Search 
target list. Ca\,\II\ activity indices, which are essential
in the interpretation of planet detection claims, have been determined for these stars and
placed on the Mount Wilson \rhk\ system.\\

\end{abstract}


\section{Introduction}
\label{intro}

Extensive planet search programmes are now operating on a variety of
telescopes around the world, including the 3.9m Anglo-Australian Telescope 
(Tinney et al. 2001,Butler et al. 2001),
the 10m Keck I (Vogt et al. 2000), the Lick 3m (Fischer et al. 2001), 
the Leonard Euler 1.2m (Naef 2001) and ESO Coud\'e Auxiliary Telescopes (K\"urster 2000)
at La Silla, and the Observatoire 
de Haute-Provence 1.9m (Queloz et al. 1998). These programmes reach velocity precisions 
ranging from 15 to 3\,\ms. At these levels {\em apparent} 
velocity variability (known as jitter), which is intrinsic to the target
star itself (rather than the presence of an unseen companion),
can be produced by the effects of stellar activity and rapid rotation combined with surface inhomogeneity  (Saar et al.
1998, Saar \& Fischer 2000, Santos et al. 2000).\\

\begin{figure*}
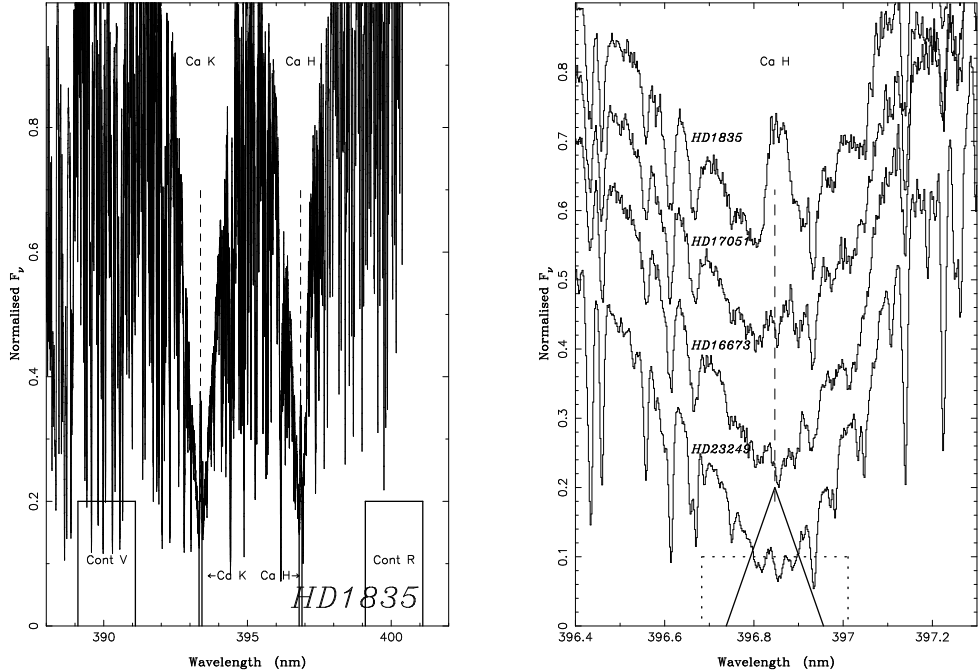

 \centerline{\psfig{file=figure1a.ps,width=2.3in}\qquad\qquad\psfig{file=figure1b.ps,width=2.3in} }
 \caption{Example Ca\,\II\ HK spectra from UCLES. The left panel shows an observation of HD1835, indicating the locations of the Ca\,\II\ H\&K lines, and the continuum V and R regions, described in Section \ref{sindex}. The fluctuations seen in this spectrum are
due to weak absorption lines, not noise. The right panel shows an expanded region of the Ca\,\II\ H line for an activity sequence from HD1835, HD16673, HD17051 to HD23429 (from top to bottom -- see Table \ref{mytable}). Also shown are the Mount Wilson triangular
Ca\,\II\ bandpass we have used{\em (solid line)}, and the Henry et al. (1996) broad Ca\,\II\ bandpass {\em (dotted line)}. All spectra have been normalised to F$_\nu$=1.0 in the R continuum region. The spectra have been offset by 0.4,0.1,0.2,0.0 (from top to bottom) for clarity.}
 \label{spectra}
\end{figure*}

The level of stellar activity in solar-type stars is commonly
examined by measuring the strength  of chromospheric emission features in the
cores of the Ca\,\II\ H and K absorption lines. This emission 
is usually parametrised by the Mount Wilson HK Project's \rhk\ 
index (see Baliunas et al. (1998) and references therein). 
This index has been shown to be a useful indicator 
of the levels of jitter in F-M type stars (Saar, Butler \& Marcy 1998). 
Estimates of \rhk\ are therefore
essential to planet search programmes, since they indicate the level 
of velocity variability
in a given star which could be erroneously
ascribed to the presence of an unseen planet,
rather than jitter. \\

Unfortunately, while extensive \rhk\ data are available in the 
Northern Hemisphere from the Mount Wilson HK Project programme 
(Baliunas et al. 1998) and the Vienna-KPNO Ca\,\II\ H\&K Survey 
(Strassmeier et al. 2000), the only large \rhk\ data set available
in the South is that of Henry et al. (1996). Henry et al. measured 
chromospheric Ca\,\II\ emission
in more than 800 southern stars (mostly G dwarfs). However, a large
fraction of the stars targeted by the Anglo-Australian Planet 
Search (Tinney et al. 2001, Butler et al. 2001) were not observed by 
Henry et al. We have therefore begun a programme of ultraviolet 
(UV) spectroscopy of these stars on the
AAT to fill this gap.\\

\section{Observations}
\label{observations}
The Anglo-Australian Planet Search (AAPS) is being carried out on the 
3.9\,m Anglo-Australian Telescope (AAT), using the 
University College London Echelle Spectrograph (UCLES). 
The observations reported
here were obtained on 2001 August 4. For these observations
UCLES was operated in its 31\,lines\,mm$^{-1}$  mode, with its UV-optimised
collimator, to provide an echellogram centered at 390\,nm. 
The detector  used was the AAO's EEV 2048$\times$4096 13.5$\mu$m pixel 
detector (rather than the MITLL2 2048$\times$4096 15$\mu$m pixel CCD 
used for all previous AAPS observations). This CCD
provides excellent UV response, with a quantum efficiency 
at the Ca\,\II\ HK lines of $\approx$ 65 per cent. Compared to the MITLL2
the EEV has smaller pixels and reduced charge diffusion,
resulting in greatly improved resolution -- it will become 
the default detector for all future AAPS observing. CCD gain was 
2.7\,$e^-$\,adu$^{-1}$, and read noise was 4.4\,$e^-$ per pixel. The CCD was
binned by two spatially, to provide an effective slit length of 
11 pixels. In the configuration adopted the dispersion at the Ca\,\II\ \rhk\
lines is 0.002\,nm\,pixel$^{-1}$, while the resolution is 3.5 pixels or 0.007\,nm.\\

Sixty seven F, G, K and M-type stars were observed in 0{\farcs}7 seeing
though a 1{\farcs}0$\times$3{\farcs}5 slit, 
with exposure times ranging from 30s to 180s, providing a signal-to-noise ratio
per 0.002\,nm wavelength pixel of between 10 and 50. The stars observed 
included AAPS target stars which were also
observed by Henry et al. (1996), and stars from the Mount 
Wilson HK Project, which Henry et al. used as standards for 
calibrating their observations onto the Mount Wilson \rhk\ system. 
The stars observed are listed in Table \ref{mytable} along
with magnitude, colour and spectral type data for each 
star from the HIPPARCOS catalogue (ESA 1997).\\

\begin{table*}
  \center
  \caption{AAPS Ca\,\II\ \rhk\ Targets.}
  \begin{tabular}{llcclccccccc}
& Star      & V    & B-V & Sp.Typ.      
    &S$_{AAT}^a$ & S$_{Henry}^b$& S$_{MW}^b$ & {log\,\rhkAAT} &
    {log\,\rhkCTIO}$^c$& {log\,\rhkMW}$^c$\\[5pt]
Calibrators &
   HD 1835 & 6.39 & 0.66 & G3V    &0.319$\pm$0.001 & 0.343 & 0.347 & -4.49 & -4.44 & -4.44\\
&  HD 3795 & 6.14 & 0.72 & G3/G5V &0.153$\pm$0.001 & 0.146 & 0.156 & -5.06 & -5.11 & -5.04\\
&  HD 7570 & 4.97 & 0.57 & F8V    &0.161$\pm$0.001 & 0.161 &  -    & -4.91 & -4.95 &  -   \\
& HD 10180 & 7.33 & 0.63 & G1V    &0.161$\pm$0.002 & 0.153 &  -    & -4.98 & -5.04 &  -   \\
& HD 10700 & 3.49 & 0.73 & G8V    &0.163$\pm$0.000 & 0.172 & 0.171 & -5.00 & -4.96 & -4.96\\
& HD 11131 & 6.72 & 0.65 & G0     &0.311$\pm$0.001 & 0.310 & 0.336 & -4.50 & -4.47 & -4.43\\
& HD 13445 & 6.12 & 0.81 & K0V    &0.297$\pm$0.001 & 0.251 &  -    & -4.65 & -4.74 &  -   \\
& HD 16673 & 5.79 & 0.52 & F6V    &0.209$\pm$0.001 & 0.225 & 0.215 & -4.69 & -4.62 & -4.66\\
& HD 17051 & 5.40 & 0.56 & G0V    &0.240$\pm$0.001 & 0.225 &  -    & -4.61 & -4.65 &  -   \\
& HD 18907 & 5.88 & 0.79 & G8/K0V &0.157$\pm$0.001 & 0.109 &  -    & -5.05 & -5.40 &  -   \\
& HD 20201 & 7.27 & 0.58 & G0V    &0.174$\pm$0.001 & 0.164 &  -    & -4.88 & -4.93 &  -   \\
& HD 20766 & 5.53 & 0.64 & G2V    &0.262$\pm$0.001 & 0.245 &  -    & -4.60 & -4.65 &  -   \\
& HD 20782 & 7.36 & 0.63 & G3V    &0.180$\pm$0.001 & 0.176 &  -    & -4.88 & -4.91 &  -   \\
& HD 20807 & 5.24 & 0.60 & G1V    &0.176$\pm$0.001 & 0.195 &  -    & -4.88 & -4.76 &  -   \\
& HD 23079 & 7.12 & 0.58 & F8/G0V &0.155$\pm$0.001 & 0.164 &  -    & -4.99 & -4.94 &  -   \\
& HD 23249 & 3.52 & 0.92 & K0IV   &0.141$\pm$0.000 & 0.129 & 0.137 & -5.17 & -5.22 & -5.18\\
& HD 25874 & 6.74 & 0.67 & G5IV-V &0.164$\pm$0.001 & 0.173 &  -    & -4.96 & -4.93 &  -   \\
& HD 38283 & 6.69 & 0.58 & G0/G1V &0.153$\pm$0.001 & 0.158 &  -    & -5.01 & -4.97 &  -   \\
& HD 38973 & 6.63 & 0.59 & G2V    &0.158$\pm$0.002 & 0.175 &  -    & -5.01 & -4.88 &  -   \\
&HD 202628 & 6.75 & 0.64 & G5V    &0.250$\pm$0.001 & 0.217 &  -    & -4.63 & -4.73 &  -   \\
&HD 205536 & 7.07 & 0.76 & G8V    &0.162$\pm$0.001 & 0.179 &  -    & -5.01 & -4.94 &  -   \\
&HD 212168 & 6.12 & 0.60 & G3IV  &0.157$\pm$0.001 & 0.173 &  -    & -4.99 & -4.89 &  -   \\
&HD 222335 & 7.18 & 0.80 & K1V    &0.233$\pm$0.003 & 0.256 &  -    & -4.78 & -4.73 &  -   \\[5pt]
New Data
&   HD 142 & 5.70 & 0.52 & G1IV  &0.155$\pm$0.001 &  -    &  -    & -4.95 &  -    &  -   \\
&  HD 2039 & 9.00 & 0.66 & G4V   &0.176$\pm$0.005 &  -    &  -    & -4.91 &  -    &  -   \\
&  HD 2587 & 8.46 & 0.75 & G6V   &0.151$\pm$0.003 &  -    &  -    & -5.08 &  -    &  -   \\
&  HD 3823 & 5.89 & 0.56 & G1V   &0.155$\pm$0.001 &  -    &  -    & -4.98 &  -    &  -   \\
&  HD 6735 & 7.01 & 0.57 & F8V   &0.166$\pm$0.002 &  -    &  -    & -4.91 &  -    &  -   \\
&  HD 7199 & 8.06 & 0.85 & K0IV/V&0.180$\pm$0.004 &  -    &  -    & -4.98 &  -    &  -   \\
&  HD 9280 & 8.03 & 0.76 & G5    &0.150$\pm$0.003 &  -    &  -    & -5.08 &  -    &  -   \\
& HD 10647 & 5.52 & 0.55 & F8V   &0.204$\pm$0.001 &  -    &  -    & -4.72 &  -    &  -   \\
& HD 11112 & 7.13 & 0.64 & G4V   &0.156$\pm$0.001 &  -    &  -    & -5.02 &  -    &  -   \\
& HD 16417 & 5.78 & 0.65 & G1V   &0.147$\pm$0.001 &  -    &  -    & -5.09 &  -    &  -   \\
& HD 19632 & 7.29 & 0.68 & G3/G5V&0.362$\pm$0.002 &  -    &  -    & -4.42 &  -    &  -   \\
& HD 20029 & 7.05 & 0.56 & F7V   &0.149$\pm$0.001 &  -    &  -    & -5.02 &  -    &  -   \\
& HD 22104 & 8.32 & 0.68 & G3V   &0.154$\pm$0.004 &  -    &  -    & -5.04 &  -    &  -   \\
& HD 23127 & 8.58 & 0.69 & G2V   &0.163$\pm$0.004 &  -    &  -    & -4.99 &  -    &  -   \\
& HD 23484 & 6.99 & 0.87 & K1V   &0.510$\pm$0.002 &  -    &  -    & -4.43 &  -    &  -   \\
& HD 24112 & 7.24 & 0.56 & F8V   &0.152$\pm$0.001 &  -    &  -    & -5.00 &  -    &  -   \\
& HD 25587 & 7.40 & 0.54 & F7V   &0.158$\pm$0.002 &  -    &  -    & -4.95 &  -    &  -   \\
& HD 26754 & 7.16 & 0.55 & F7/F8V&0.152$\pm$0.001 &  -    &  -    & -5.00 &  -    &  -   \\
& HD 27442 & 4.44 & 1.08 & K2IV  &0.150$\pm$0.001 &  -    &  -    & -5.27 &  -    &  -   \\
& HD 28255 & 6.28 & 0.66 & G4V   &0.181$\pm$0.001 &  -    &  -    & -4.89 &  -    &  -   \\
&GJ9155B$^d$
          & 7.50 & 0.71 & G6V   &0.258$\pm$0.001 &  -    &  -    & -4.65 &  -    &  -   \\
& HD 30177 & 8.41 & 0.77 & G8V   &0.151$\pm$0.004 &  -    &  -    & -5.08 &  -    &  -   \\
& HD 30876 & 7.49 & 0.90 & K2V   &0.469$\pm$0.002 &  -    &  -    & -4.51 &  -    &  -   \\
& HD 31527 & 7.49 & 0.61 & G2V   &0.165$\pm$0.001 &  -    &  -    & -4.95 &  -    &  -   \\
& HD 36108 & 6.78 & 0.59 & G3V   &0.153$\pm$0.002 &  -    &  -    & -5.01 &  -    &  -   \\
& HD 38382 & 6.34 & 0.58 & F8/G0V&0.162$\pm$0.001 &  -    &  -    & -4.94 &  -    &  -   \\
& HD 40307 & 7.17 & 0.94 & K3V   &0.267$\pm$0.004 &  -    &  -    & -4.83 &  -    &  -   \\
&HD 204385 & 7.14 & 0.56 & G0IV  &0.158$\pm$0.001 &  -    &  -    & -4.98 &  -    &  -   \\
&HD 204961 & 8.66 & 1.52 & M1V   &0.975$\pm$0.008 &  -    &  -    & -5.10 &  -    &  -   \\
&HD 205390 & 7.14 & 0.88 & K2V   &0.424$\pm$0.002 &  -    &  -    & -4.53 &  -    &  -   \\
&HD 209268 & 6.88 & 0.56 & F7V   &0.153$\pm$0.001 &  -    &  -    & -4.99 &  -    &  -   \\
&HD 211317 & 7.26 & 0.65 & G4IV  &0.156$\pm$0.002 &  -    &  -    & -5.03 &  -    &  -   \\
&HD 216437 & 6.04 & 0.66 & G4IV-V&0.154$\pm$0.001 &  -    &  -    & -5.04 &  -    &  -   \\
&HD 217987 & 7.34 & 1.48 & M2/M3V&1.074$\pm$0.004 &  -    &  -    & -5.01 &  -    &  -   \\
&HD 222237 & 7.09 & 0.99 & K3V   &0.281$\pm$0.007 &  -    &  -    & -4.86 &  -    &  -   \\
   \end{tabular}               
                               
\raggedright                   
\noindent                      
Notes : 
$a$ - S$_{\mathrm AAT}$ calculated as described in Section \ref{analysis}, {\em after}
      calibration onto the Mount Wilson S$_{\mathrm MW}$ index. 
$b$ - S$_{Henry}$ is the Ca\,\II\ HK index of Henry et al. (1996) (calibrated onto the
      same scale as the Mount Wilson S$_{MW}$ index), as provided in columns 8-11
      of their Table 3, and column 14 of their Table 4.  S$_{MW}$ is the Mount Wilson
      index used to perform this calibration from columns 4-7 of their Table 3.
$c$ - log\,\rhkCTIO\ are the log\,\rhk\ indices measured by Henry et al. (1996), while
      log\,\rhkMW\ are the log\,\rhk\ indices Henry et al. used to check their 
      calibration.
$d$ - Also known as SAO258997. This star lies within 6" of 
      HD28255 (itself also known as GJ9155A) 
\label{mytable}
\end{table*}

\section{Analysis}
\label{analysis}

\subsection{Image Processing}

Reduction of the CCD echellogram images proceeded in a standard manner
in the Figaro environment. Images were overscan subtracted, trimmed and
then flat-fielded using flat-field exposures from which the spectrograph
blaze function for each order had been fitted and divided out. Scattered
light in the spectrograph produces a faint overall level of illumination
of the detector -- this was fitted in the
gaps between orders and subtracted over the whole detector. These images
were then straightened to make the echelle orders align with columns on the
detector. The entire 3{\farcs}5 slit length was then extracted from each order --
no sky subtraction was performed. In the wavelength range of interest
the sky is featureless. An observation of blank sky obtained well
after astronomical twilight reveals sky counts contribute less 
than 0.016\,$e^-$\,s$^{-1}$ per 0.002\,nm wavelength pixel, which
is negligible compared to our targets 
($\approx$8\,$e^-$\,s$^{-1}$ per 0.002\,nm wavelength pixel at V=7.5).

The orders containing the Ca\,\II\ H and K lines, as well as the adjoining
orders (orders 146-142), were then extracted for individual processing. All
were wavelength calibrated using a ThAr spectrum obtained at the 
end of the night. All these data were also flux calibrated using an 
observation of the HST flux standard $\mu$\,Col (Turnshek et al. 1990) -- note
that we use this observation only to correct the blaze and inter-order
sensitivity of our data. The night was not spectro-photometric.

The spectra in order 142 ($\lambda = 398.0-402.6$\,nm) were
cross-correlated against the observation of HD142, and the 
barycentric correction for HD142 was applied. The result is a
set of zero-velocity spectra. (Though HD142 is a radial velocity variable at
the 30\ms\ level (Tinney et al. 2002), this variability is negligible for our purposes).
Lastly, all the spectra
were re-sampled to the same 0.002\,nm binning.

\subsection{S Indices}
\label{sindex}

The resulting spectra were then used to produce a 
chromospheric Ca\,\II\ HK emission index, S$_{\mathrm MW}$. 
The Mount Wilson HK Project
defines its S$_{\mathrm MW}$  (Duncan et al. 1991) as the ratio of the flux in 
two triangular bands with
full-width at half-maximum of 0.109\,nm centered at 393.3664 (K) and 
396.8470\,nm (H), to that in two 2.0\,nm wide
continuum bands centered at 390.1 (V) and 400.1\,nm (R). This
ratio is measured using a specifically-designed multi-channel photometer. S$_{\mathrm MW}$
defined as 
$$ {\mathrm S_{\mathrm MW}} = \alpha \frac{N_H + N_K}{N_R + N_V}, $$
where the $N_i$ are the counts in each of the H, K, V and R bands, and $\alpha$
is a constant determined for each night by the observation
of standards.

\begin{figure}
 \centerline{\psfig{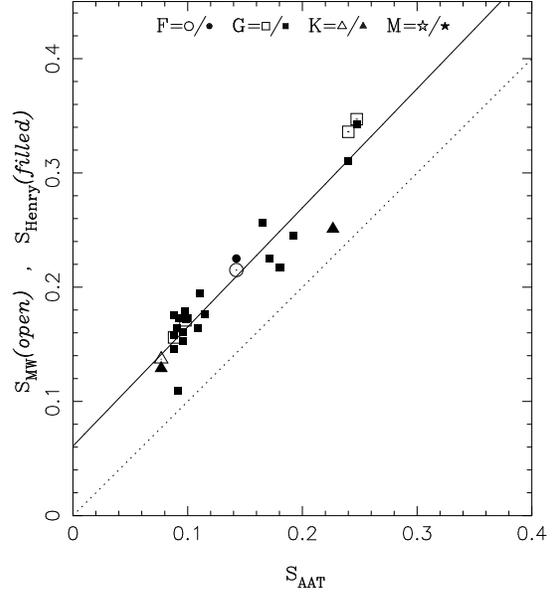} }
 \caption{Calibration of the S$_{\mathrm AAT}$ spectral index on to
 the S index scale of Mount Wilson {\em (open symbols)} and Henry et al. (1996) 
 {\em (filled symbols)} surveys, for
 the observed calibration objects. Different spectral types are plotted with different
 symbols. The dotted line shows a unitary
 transformation, and the solid line is the fitted linear calibration
  (S$_{\mathrm MW}$ = 1.043\,S$_{\mathrm AAT}$  + 0.061).}
 \label{scal}
\end{figure}

Using the high-resolution spectra obtained at the AAT, we were able to define
our Ca\,\II\ H and K and continuum bands as having the same effective
profiles (i.e. 0.1\,nm triangular and 2.0\,nm square bandpasses)
as used at Mount Wilson. At the AAT, therefore, we measure the
flux in each band, and normalise for the band width to obtain
$$N = \frac{\sum_\lambda F(\lambda) \Delta\lambda}{\sum_\lambda \Delta\lambda} $$
in each band. Because both the Ca H and Ca K lines appear in two orders
in our echellogram, we actually obtain two flux estimates for each, which
we combine to form an observed S$_{\mathrm AAT}$ index as follows
$$ {\mathrm S_{\mathrm AAT}} = \frac{N_{H,1} + N_{H,2} + N_{K,1} + N_{K,2}} 
                               {2 ( N_R + N_V)}. $$
In Figure \ref{scal} we plot this
observed S$_{\mathrm AAT}$ against the S$_{\mathrm MW}$ and S$_{\mathrm Henry}$
values reported by Henry et al. A tight linear correlation is
seen between S$_{\mathrm MW}$ and S$_{\mathrm AAT}$, with rms scatter about 
a linear fit of 0.02, zero-point offset of 0.061 and slope of 1.043.
Examination of  Fig. \ref{scal} would suggest that a calibration against
the MW data alone would have even smaller scatter, and could be preferred.
Certainly our high-resolution spectra should  transfer onto the
MW system in a linear fashion with no systematic effects. Unfortunately,
the number of MW stars accessible in the south is small, so such
a calibration would (at present) be based on a small number of stars, 
and only a few active ones. Because such active stars are by their
very nature highly variable in their activity, we have therefore chosen 
to use {\em all} the MW and Henry et al. calibrators to derive the relation shown in 
Fig. \ref{scal}.
No major correlation is seen between degree of deviation from
the fit and spectral type. 

\begin{figure}
 \centerline{\psfig{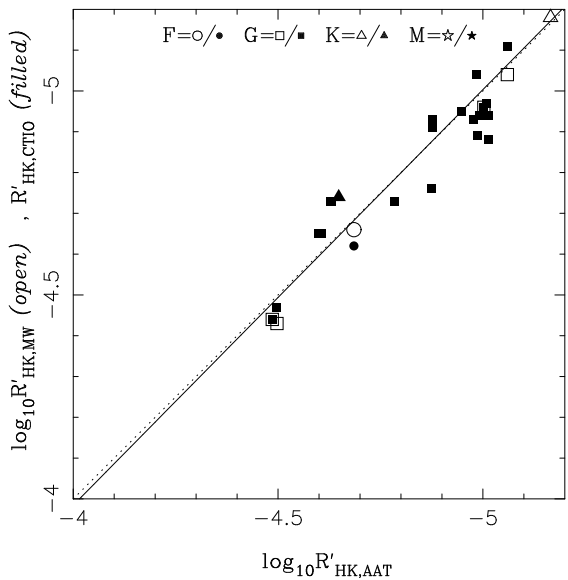} }
 \caption{Plot of the derived \rhkAAT\ versus the  
 Mount Wilson {\em (open symbols)} and Henry et al. (1996) {\em (filled symbols)}
 \rhk\ measurements for our 
 calibration objects. The dotted line shows a unitary
 transformation, and the solid line is a linear fit.}
 \label{rhk}
\end{figure}

The values of S$_{\mathrm AAT}$ so calibrated (including propagated photon-counting
uncertainties), and the calibrating S indices are shown in Table \ref{mytable}.
In general the scatter about the calibration curve is significantly larger
than the photon-counting uncertainties, which likely reflects the intrinsically
variable nature of stellar activity.

It should be noted that an advantage of performing these observations at
high spectral resolution is that we can define our bandpasses in exactly the same
way as those used at Mount Wilson -- the resulting indices require only a  linear calibration  with slope of 1.0 to place them on a standard system.  Henry et al., in contrast, were forced to
use a non-linear calibration, which changes from run to run, introducing possible
systematic effects. Experiments on our data show that when the Ca\,\II\ bandpass 
is broadened, non-linear calibrations of increasing complexity become necessary, 
with the consequent risk of introducing systematic uncertainties.

\subsection{\rhk\ Indices}

The S index provides an estimator of photospheric {\em plus}
chromospheric Ca\,\II\ strength. For activity purposes, however,
it is the chromospheric emission which is of interest.
This is usually normalised to the star's bolometric
luminosity, and parametrised by the \rhk\ index. 
The derivation of \rhk\ 
 from S is described in detail in the
Appendix of Noyes et al. (1984). We follow these procedures to
derive the 
log\,\rhkAAT\ values shown in Table \ref{mytable}, where we also show:
the log\,\rhkCTIO\ values measured by Henry et al. (1996);
the log\,\rhkMW\ values used to calibrate those data; and the HIPPARCOS
B--V photometry used to estimate the photometric contribution to
S$_{\mathrm AAT}$.

In Figure \ref{rhk} we plot log\,\rhkAAT\ against log\,\rhkMW\ and log\,\rhkCTIO.
Note that there are no arbitrary
calibration factors included in the transformation from S$_{\mathrm AAT}$ to
\rhkAAT. The resulting plot is consistent with a unitary transformation -- the
solid line fit shown in Fig. \ref{rhk} is for a zero-offset of 0.09$\pm$2.3 and
a slope 1.02$\pm$0.09, and has rms scatter of 0.09 (for all data), or 0.05
for the MW data alone . We therefore apply no 
further calibration to our log\,\rhkAAT\
estimates, presented in Table \ref{mytable}, but note a root-mean-square
scatter about the 
MW standards of $\pm$0.05 (in the logarithm), which is a better
uncertainty estimate for the final results than the 
photon-counting uncertainties. The rms scatter about the fit for observations
in common with Henry et al. is larger at $\pm$0.09. Expressed as the ``mean absolute
deviation'' (used by Henry et al.) this corresponds to $\pm$0.07, 
while the mean absolute deviation  for 
the calibration of the Henry et al. sample onto the
MW data is $\pm$0.052, leading us to conclude a large part
of this scatter may be intrinsic to the Henry et al.
comparison sample rather than the AAT data.

\section{Discussion}
\label{discussion}

Table \ref{mytable} provides at least two new \rhk\ measurements for 
stars with known planetary companions, HD27442 (Butler et al. 2001)
and HD142 (Tinney et al. 2002) --
though there may be several more, as yet undiscovered, planets in this
set of stars (for which AAPS observing is on-going).
In the case of HD27442, we find the star, at this epoch, to be very inactive.
Saar et al. (1998) and Santos et al. (2000) 
have examined the levels of velocity variability
intrinsic to stars themselves ($\sigma^\prime_v$), as a function of 
activity (\rhk), in stars 
monitored by the Lick 3m planet search.
A trend emerges in which more active stars show more intrinsic velocity
instability, or ``jitter'', with a form of 
$\sigma^\prime_v \propto $(\rhk)$^{1.1}$
for G and K dwarfs, and with $\sigma^\prime_v$ in the range 3-10\,\ms\ for
log\,\rhk=-5.0.

Based on this we would predict the jitter in HD27442 to be $\lid$1.5-5\,\ms.
Butler et al. (2001) derive a scatter about their Keplerian fit to
the velocity data for HD27442 of 2.9\,\ms, which is an upper limit to
the level of jitter which could be present in this star -- in line with
its extremely low level of Ca\,\II\ activity.

HD142 is only slightly more active at log\,\rhk =-4.95, implying an expected
level of jitter in the range $\lid$4-12\,\ms. Tinney et al. (2002) report
a scatter about a single Keplerian fit of 5.9\,\ms, consistent with this prediction.

Several of the stars in Table \ref{mytable} reveal themselves to be quite
active, with log\,\rhk $>$ -4.5. This would correspond to likely levels of jitter 
greater than 10-30\,m/s in HD19632, HD30876 and HD205390.
Planet candidates detected around stars like this must be scrutinised with
great care to avoid erroneous detection claims. The recent results
of Queloz et al. (2001) for HD166435, in particular, provide a valuable
cautionary tale. HD166435 was selected for observation as part of the Observatoire
de Haute-Provence ELODIE planet search. The available data (which did not include
\rhk\ measurements) did not indicate particular youth or activity. Over the 
period May 1997 to September 1999 seventy spectra of this star were acquired,
as it appeared to have an unseen planetary companion. Further analysis, however,
revealed this variability to be entirely due to activity. A result which is
not unexpected when the value of the S index of this star 
(as measured at Mount Wilson) is considered. At S=0.42-0.51 (Queloz et al. 2002)
this G0 dwarf is phenomenally active -- more active than the most active stars
shown in Figs. \ref{scal} and \ref{rhk}, implying log\,\rhk $\approx -4$. At this level
of activity the Saar et al. (1998) results would imply an expected level of
jitter in the range 36-120\,\ms, which would go much of the way toward explaining
the 200\,\ms\ amplitude velocity variations seen in HD164435. Having \rhk\ measurements
in hand when considering the validity of a planet candidate is clearly essential.

Finally we would note the intrinsically variable nature of
active stars, and the importance of obtaining Ca\,\II\ observations at multiple epochs
in order to be sure a star really {\em is} inactive. The reliability
of such data is important
when trying to decide what level of velocity variability could be due
to activity/rotation induced jitter, and what level can be trusted to be
due to an unseen companion.

\section{Conclusion}
\label{conclusion}

Ca\,\II\ HK activity measurements are provided for 59 stars from the
Anglo-Australian Planet Search target list -- thirty five of which are for stars
not previously published. Four-metre telescope UV echelle spectroscopy
is an efficient way in which to obtain high quality data for this purpose, and
because of the high resolution of such data, calibration onto the
standard Mount Wilson \rhk\ system can be carried out with greatly
reduced systematic uncertainties.

\subsection*{Acknowledgments}
\label{Acknowledgments}

The Anglo-Australian Planet Search team would like to gratefully 
acknowledge the grant of Director's time in which these observations
were carried out, the superb technical support which has been
received throughout the programme from AAT staff - in particular E.Penny, 
R.Patterson, D.Stafford, F.Freeman, S.Lee, J.Pogson and G.Schafer. 
We further acknowledge support 
by; the partners of the Anglo-Australian Telescope Agreement (CGT,HRAJ,APJ);
NASA grant NAG5-8299 \& NSF grant AST95-20443 (GWM);
NSF grant AST-9988087 (RPB); and Sun Microsystems.

\bsp

\label{lastpage}

\end{document}

%% file: psfig2.tex
\def\PsfigVersion{1.9}
\ifx\undefined\psfig\else \fi

\let\LaTeXAtSign=\@
\let\@=\relax
\edef\psfigRestoreAt{\catcode`\@=\number\catcode`@\relax}
\catcode`\@=11\relax
\newwrite\@unused
\def\ps@typeout#1{{\let\protect\string\immediate\write\@unused{#1}}}
\ps@typeout{psfig/tex \PsfigVersion}

\def\figurepath{./}

\def\@nnil{\@nil}
\def\@empty{}
\def\@psdonoop#1\@@#2#3{}
\def\@psdo#1:=#2\do#3{\edef\@psdotmp{#2}\ifx\@psdotmp\@empty \else
    \expandafter\@psdoloop#2,\@nil,\@nil\@@#1{#3}\fi}
\def\@psdoloop#1,#2,#3\@@#4#5{\def#4{#1}\ifx #4\@nnil \else
       #5\def#4{#2}\ifx #4\@nnil \else#5\@ipsdoloop #3\@@#4{#5}\fi\fi}
\def\@ipsdoloop#1,#2\@@#3#4{\def#3{#1}\ifx #3\@nnil 
       \let\@nextwhile=\@psdonoop \else
      #4\relax\let\@nextwhile=\@ipsdoloop\fi\@nextwhile#2\@@#3{#4}}
\def\@tpsdo#1:=#2\do#3{\xdef\@psdotmp{#2}\ifx\@psdotmp\@empty \else
    \@tpsdoloop#2\@nil\@nil\@@#1{#3}\fi}
\def\@tpsdoloop#1#2\@@#3#4{\def#3{#1}\ifx #3\@nnil 
       \let\@nextwhile=\@psdonoop \else
      #4\relax\let\@nextwhile=\@tpsdoloop\fi\@nextwhile#2\@@#3{#4}}
\ifx\undefined\fbox
\newdimen\fboxrule
\newdimen\fboxsep
\newdimen\ps@tempdima
\newbox\ps@tempboxa
\long\def\fbox#1{\leavevmode\setbox\ps@tempboxa\hbox{#1}\ps@tempdima\fboxrule
    \advance\ps@tempdima \fboxsep \advance\ps@tempdima \dp\ps@tempboxa
   \hbox{\lower \ps@tempdima\hbox
  {\vbox{\hrule height \fboxrule
          \hbox{\vrule width \fboxrule \hskip\fboxsep
          \vbox{\vskip\fboxsep \box\ps@tempboxa\vskip\fboxsep}\hskip 
                 \fboxsep\vrule width \fboxrule}
                 \hrule height \fboxrule}}}}
\fi
\newread\ps@stream
\newif\ifnot@eof       
\newif\if@noisy        
\newif\if@atend        
\newif\if@psfile       
{\catcode`\%=12\global\gdef\epsf@start{
\def\epsf@PS{PS}
\def\epsf@getbb#1{%
\openin\ps@stream=#1
\ifeof\ps@stream\ps@typeout{Error, File #1 not found}\else
   {\not@eoftrue \chardef\other=12
    \def\do##1{\catcode`##1=\other}\dospecials \catcode`\ =10
    \loop
       \if@psfile
	  \read\ps@stream to \epsf@fileline
       \else{
	  \obeyspaces
          \read\ps@stream to \epsf@tmp\global\let\epsf@fileline\epsf@tmp}
       \fi
       \ifeof\ps@stream\not@eoffalse\else
       \if@psfile\else
       \expandafter\epsf@test\epsf@fileline:. \\%
       \fi
          \expandafter\epsf@aux\epsf@fileline:. \\%
       \fi
   \ifnot@eof\repeat
   }\closein\ps@stream\fi}%
\long\def\epsf@test#1#2#3:#4\\{\def\epsf@testit{#1#2}
			\ifx\epsf@testit\epsf@start\else
\ps@typeout{Warning! File does not start with `\epsf@start'.  It may not be a PostScript file.}
			\fi
			\@psfiletrue} 
{\catcode`\%=12\global\let\epsf@percent=
\long\def\epsf@aux#1#2:#3\\{\ifx#1\epsf@percent
   \def\epsf@testit{#2}\ifx\epsf@testit\epsf@bblit
	\@atendfalse
        \epsf@atend #3 . \\%
	\if@atend	
	   \if@verbose{
		\ps@typeout{psfig: found `(atend)'; continuing search}
	   }\fi
        \else
        \epsf@grab #3 . . . \\%
        \not@eoffalse
        \global\no@bbfalse
        \fi
   \fi\fi}%
\def\epsf@grab #1 #2 #3 #4 #5\\{%
   \global\def\epsf@llx{#1}\ifx\epsf@llx\empty
      \epsf@grab #2 #3 #4 #5 .\\\else
   \global\def\epsf@lly{#2}%
   \global\def\epsf@urx{#3}\global\def\epsf@ury{#4}\fi}%
\def\epsf@atendlit{(atend)} 
\def\epsf@atend #1 #2 #3\\{%
   \def\epsf@tmp{#1}\ifx\epsf@tmp\empty
      \epsf@atend #2 #3 .\\\else
   \ifx\epsf@tmp\epsf@atendlit\@atendtrue\fi\fi}

\chardef\psletter = 11 
\chardef\other = 12

\newif \ifdebug 
\newif\ifc@mpute 
\c@mputetrue 

\let\then = \relax
\def\r@dian{pt }
\let\r@dians = \r@dian
\let\dimensionless@nit = \r@dian
\let\dimensionless@nits = \dimensionless@nit
\def\internal@nit{sp }
\let\internal@nits = \internal@nit
\newif\ifstillc@nverging
\def \Mess@ge #1{\ifdebug \then \message {#1} \fi}

{ 
	\catcode `\@ = \psletter
	\gdef \nodimen {\expandafter \n@dimen \the \dimen}
	\gdef \term #1 #2 #3%
	       {\edef \t@ {\the #1}
		\edef \t@@ {\expandafter \n@dimen \the #2\r@dian}%
		\t@rm {\t@} {\t@@} {#3}%
	       }
	\gdef \t@rm #1 #2 #3%
	       {{%
		\count 0 = 0
		\dimen 0 = 1 \dimensionless@nit
		\dimen 2 = #2\relax
		\Mess@ge {Calculating term #1 of \nodimen 2}%
		\loop
		\ifnum	\count 0 < #1
		\then	\advance \count 0 by 1
			\Mess@ge {Iteration \the \count 0 \space}%
			\Multiply \dimen 0 by {\dimen 2}%
			\Mess@ge {After multiplication, term = \nodimen 0}%
			\Divide \dimen 0 by {\count 0}%
			\Mess@ge {After division, term = \nodimen 0}%
		\repeat
		\Mess@ge {Final value for term #1 of 
				\nodimen 2 \space is \nodimen 0}%
		\xdef \Term {#3 = \nodimen 0 \r@dians}%
		\aftergroup \Term
	       }}
	\catcode `\p = \other
	\catcode `\t = \other
	\gdef \n@dimen #1pt{#1} 
}

\def \Divide #1by #2{\divide #1 by #2} 

\def \Multiply #1by #2
       {{
	\count 0 = #1\relax
	\count 2 = #2\relax
	\count 4 = 65536
	\Mess@ge {Before scaling, count 0 = \the \count 0 \space and
			count 2 = \the \count 2}%
	\ifnum	\count 0 > 32767 
	\then	\divide \count 0 by 4
		\divide \count 4 by 4
	\else	\ifnum	\count 0 < -32767
		\then	\divide \count 0 by 4
			\divide \count 4 by 4
		\else
		\fi
	\fi
	\ifnum	\count 2 > 32767 
	\then	\divide \count 2 by 4
		\divide \count 4 by 4
	\else	\ifnum	\count 2 < -32767
		\then	\divide \count 2 by 4
			\divide \count 4 by 4
		\else
		\fi
	\fi
	\multiply \count 0 by \count 2
	\divide \count 0 by \count 4
	\xdef \product {#1 = \the \count 0 \internal@nits}%
	\aftergroup \product
       }}

\def\r@duce{\ifdim\dimen0 > 90\r@dian \then   
		\multiply\dimen0 by -1
		\advance\dimen0 by 180\r@dian
		\r@duce
	    \else \ifdim\dimen0 < -90\r@dian \then  
		\advance\dimen0 by 360\r@dian
		\r@duce
		\fi
	    \fi}

\def\Sine#1%
       {{%
	\dimen 0 = #1 \r@dian
	\r@duce
	\ifdim\dimen0 = -90\r@dian \then
	   \dimen4 = -1\r@dian
	   \c@mputefalse
	\fi
	\ifdim\dimen0 = 90\r@dian \then
	   \dimen4 = 1\r@dian
	   \c@mputefalse
	\fi
	\ifdim\dimen0 = 0\r@dian \then
	   \dimen4 = 0\r@dian
	   \c@mputefalse
	\fi
	\ifc@mpute \then
		\divide\dimen0 by 180
		\dimen0=3.141592654\dimen0
		\dimen 2 = 3.1415926535897963\r@dian 
		\divide\dimen 2 by 2 
		\Mess@ge {Sin: calculating Sin of \nodimen 0}%
		\count 0 = 1 
		\dimen 2 = 1 \r@dian 
		\dimen 4 = 0 \r@dian 
		\loop
			\ifnum	\dimen 2 = 0 
			\then	\stillc@nvergingfalse 
			\else	\stillc@nvergingtrue
			\fi
			\ifstillc@nverging 
			\then	\term {\count 0} {\dimen 0} {\dimen 2}%
				\advance \count 0 by 2
				\count 2 = \count 0
				\divide \count 2 by 2
				\ifodd	\count 2 
				\then	\advance \dimen 4 by \dimen 2
				\else	\advance \dimen 4 by -\dimen 2
				\fi
		\repeat
	\fi		
			\xdef \sine {\nodimen 4}%
       }}

\def\Cosine#1{\ifx\sine\UnDefined\edef\Savesine{\relax}\else
		             \edef\Savesine{\sine}\fi
	{\dimen0=#1\r@dian\advance\dimen0 by 90\r@dian
	 \Sine{\nodimen 0}
	 \xdef\cosine{\sine}
	 \xdef\sine{\Savesine}}}	      

\def\psdraft{
	\def\@psdraft{0}
}
\def\psfull{
	\def\@psdraft{100}
}

\psfull

\newif\if@scalefirst
\def\psscalefirst{\@scalefirsttrue}
\def\psrotatefirst{\@scalefirstfalse}
\psrotatefirst

\newif\if@draftbox
\def\psnodraftbox{
	\@draftboxfalse
}
\def\psdraftbox{
	\@draftboxtrue
}
\@draftboxtrue

\newif\if@prologfile
\newif\if@postlogfile
\def\pssilent{
	\@noisyfalse
}
\def\psnoisy{
	\@noisytrue
}
\psnoisy
\newif\if@bbllx
\newif\if@bblly
\newif\if@bburx
\newif\if@bbury
\newif\if@height
\newif\if@width
\newif\if@rheight
\newif\if@rwidth
\newif\if@angle
\newif\if@clip
\newif\if@verbose
\newif\if@scale
\def\@p@@sclip#1{\@cliptrue}

\newif\if@decmpr

\def\@p@@sfigure#1{\def\@p@sfile{null}\def\@p@sbbfile{null}
	        \openin1=#1.bb
		\ifeof1\closein1
	        	\openin1=\figurepath#1.bb
			\ifeof1\closein1
			        \openin1=#1
				\ifeof1\closein1%
				       \openin1=\figurepath#1
					\ifeof1
					   \ps@typeout{Error, File #1 not found}
						\if@bbllx\if@bblly
				   		\if@bburx\if@bbury
			      				\def\@p@sfile{#1}%
			      				\def\@p@sbbfile{#1}%
							\@decmprfalse
				  	   	\fi\fi\fi\fi
					\else\closein1
				    		\def\@p@sfile{\figurepath#1}%
				    		\def\@p@sbbfile{\figurepath#1}%
						\@decmprfalse
	                       		\fi%
			 	\else\closein1%
					\def\@p@sfile{#1}
					\def\@p@sbbfile{#1}
					\@decmprfalse
			 	\fi
			\else
				\def\@p@sfile{\figurepath#1}
				\def\@p@sbbfile{\figurepath#1.bb}
				\@decmprtrue
			\fi
		\else
			\def\@p@sfile{#1}
			\def\@p@sbbfile{#1.bb}
			\@decmprtrue
		\fi}

\def\@p@@sfile#1{\@p@@sfigure{#1}}

\def\@p@@sbbllx#1{
		\@bbllxtrue
		\dimen100=#1
		\edef\@p@sbbllx{\number\dimen100}
}
\def\@p@@sbblly#1{
		\@bbllytrue
		\dimen100=#1
		\edef\@p@sbblly{\number\dimen100}
}
\def\@p@@sbburx#1{
		\@bburxtrue
		\dimen100=#1
		\edef\@p@sbburx{\number\dimen100}
}
\def\@p@@sbbury#1{
		\@bburytrue
		\dimen100=#1
		\edef\@p@sbbury{\number\dimen100}
}
\def\@p@@sheight#1{
		\@heighttrue
		\dimen100=#1
   		\edef\@p@sheight{\number\dimen100}
}
\def\@p@@swidth#1{
		\@widthtrue
		\dimen100=#1
		\edef\@p@swidth{\number\dimen100}
}
\def\@p@@srheight#1{
		\@rheighttrue
		\dimen100=#1
		\edef\@p@srheight{\number\dimen100}
}
\def\@p@@srwidth#1{
		\@rwidthtrue
		\dimen100=#1
		\edef\@p@srwidth{\number\dimen100}
}
\def\@p@@sangle#1{
		\@angletrue
		\edef\@p@sangle{#1} 
}
\def\@p@@srotate#1{\@p@@sangle{-#1}}
\def\@p@@sscale#1{
		\@scaletrue
		\edef\@p@sscale{#1}
}
\def\@p@@ssilent#1{ 
		\@verbosefalse
}
\def\@p@@sprolog#1{\@prologfiletrue\def\@prologfileval{#1}}
\def\@p@@spostlog#1{\@postlogfiletrue\def\@postlogfileval{#1}}
\def\@cs@name#1{\csname #1\endcsname}
\def\@setparms#1=#2,{\@cs@name{@p@@s#1}{#2}}
\def\ps@init@parms{
		\@bbllxfalse \@bbllyfalse
		\@bburxfalse \@bburyfalse
		\@heightfalse \@widthfalse
		\@rheightfalse \@rwidthfalse
		\@scalefalse
		\def\@p@sbbllx{}\def\@p@sbblly{}
		\def\@p@sbburx{}\def\@p@sbbury{}
		\def\@p@sheight{}\def\@p@swidth{}
		\def\@p@srheight{}\def\@p@srwidth{}
		\def\@p@sangle{0}
		\def\@p@sfile{} \def\@p@sbbfile{}
		\def\@p@scost{10}
		\def\@sc{}
		\@prologfilefalse
		\@postlogfilefalse
		\@clipfalse
		\if@noisy
			\@verbosetrue
		\else
			\@verbosefalse
		\fi
}
\def\parse@ps@parms#1{
	 	\@psdo\@psfiga:=#1\do
		   {\expandafter\@setparms\@psfiga,}}
\newif\ifno@bb
\def\bb@missing{
	\if@verbose{
		\ps@typeout{psfig: searching \@p@sbbfile \space  for bounding box}
	}\fi
	\no@bbtrue
	\epsf@getbb{\@p@sbbfile}
        \ifno@bb \else \bb@cull\epsf@llx\epsf@lly\epsf@urx\epsf@ury\fi
}	
\def\bb@cull#1#2#3#4{
	\dimen100=#1 bp\edef\@p@sbbllx{\number\dimen100}
	\dimen100=#2 bp\edef\@p@sbblly{\number\dimen100}
	\dimen100=#3 bp\edef\@p@sbburx{\number\dimen100}
	\dimen100=#4 bp\edef\@p@sbbury{\number\dimen100}
	\no@bbfalse
}
\newdimen\p@intvaluex
\newdimen\p@intvaluey
\def\rotate@#1#2{{\dimen0=#1 sp\dimen1=#2 sp
		  \global\p@intvaluex=\cosine\dimen0
		  \dimen3=\sine\dimen1
		  \global\advance\p@intvaluex by -\dimen3
		  \global\p@intvaluey=\sine\dimen0
		  \dimen3=\cosine\dimen1
		  \global\advance\p@intvaluey by \dimen3
		  }}
\def\compute@bb{
		\no@bbfalse
		\if@bbllx \else \no@bbtrue \fi
		\if@bblly \else \no@bbtrue \fi
		\if@bburx \else \no@bbtrue \fi
		\if@bbury \else \no@bbtrue \fi
		\ifno@bb \bb@missing \fi
		\ifno@bb \ps@typeout{FATAL ERROR: no bb supplied or found}
			\no-bb-error
		\fi
		\count203=\@p@sbburx
		\count204=\@p@sbbury
		\advance\count203 by -\@p@sbbllx
		\advance\count204 by -\@p@sbblly
		\edef\ps@bbw{\number\count203}
		\edef\ps@bbh{\number\count204}
		\if@angle 
			\Sine{\@p@sangle}\Cosine{\@p@sangle}
	        	{\dimen100=\maxdimen\xdef\r@p@sbbllx{\number\dimen100}
					    \xdef\r@p@sbblly{\number\dimen100}
			                    \xdef\r@p@sbburx{-\number\dimen100}
					    \xdef\r@p@sbbury{-\number\dimen100}}
                        \def\minmaxtest{
			   \ifnum\number\p@intvaluex<\r@p@sbbllx
			      \xdef\r@p@sbbllx{\number\p@intvaluex}\fi
			   \ifnum\number\p@intvaluex>\r@p@sbburx
			      \xdef\r@p@sbburx{\number\p@intvaluex}\fi
			   \ifnum\number\p@intvaluey<\r@p@sbblly
			      \xdef\r@p@sbblly{\number\p@intvaluey}\fi
			   \ifnum\number\p@intvaluey>\r@p@sbbury
			      \xdef\r@p@sbbury{\number\p@intvaluey}\fi
			   }
			\rotate@{\@p@sbbllx}{\@p@sbblly}
			\minmaxtest
			\rotate@{\@p@sbbllx}{\@p@sbbury}
			\minmaxtest
			\rotate@{\@p@sbburx}{\@p@sbblly}
			\minmaxtest
			\rotate@{\@p@sbburx}{\@p@sbbury}
			\minmaxtest
			\edef\@p@sbbllx{\r@p@sbbllx}\edef\@p@sbblly{\r@p@sbblly}
			\edef\@p@sbburx{\r@p@sbburx}\edef\@p@sbbury{\r@p@sbbury}
		\fi
		\count203=\@p@sbburx
		\count204=\@p@sbbury
		\advance\count203 by -\@p@sbbllx
		\advance\count204 by -\@p@sbblly
		\edef\@bbw{\number\count203}
		\edef\@bbh{\number\count204}
}
\def\in@hundreds#1#2#3{\count240=#2 \count241=#3
		     \count100=\count240	
		     \divide\count100 by \count241
		     \count101=\count100
		     \multiply\count101 by \count241
		     \advance\count240 by -\count101
		     \multiply\count240 by 10
		     \count101=\count240	
		     \divide\count101 by \count241
		     \count102=\count101
		     \multiply\count102 by \count241
		     \advance\count240 by -\count102
		     \multiply\count240 by 10
		     \count102=\count240	
		     \divide\count102 by \count241
		     \count200=#1\count205=0
		     \count201=\count200
			\multiply\count201 by \count100
		 	\advance\count205 by \count201
		     \count201=\count200
			\divide\count201 by 10
			\multiply\count201 by \count101
			\advance\count205 by \count201
		     \count201=\count200
			\divide\count201 by 100
			\multiply\count201 by \count102
			\advance\count205 by \count201
		     \edef\@result{\number\count205}
}
\def\ps@scaleinhundreds#1{
		\in@hundreds{#1}{\@p@sscale}{100}
		\edef#1{\@result}
}
\def\compute@wfromh{
		\in@hundreds{\@p@sheight}{\@bbw}{\@bbh}
		\edef\@p@swidth{\@result}
}
\def\compute@hfromw{
	        \in@hundreds{\@p@swidth}{\@bbh}{\@bbw}
		\edef\@p@sheight{\@result}
}
\def\compute@handw{
		\if@height 
			\if@width
			\else
				\compute@wfromh
			\fi
		\else 
			\if@width
				\compute@hfromw
			\else
				\edef\@p@sheight{\@bbh}
				\edef\@p@swidth{\@bbw}
			\fi
		\fi
}
\def\compute@resv{
		\if@rheight \else \edef\@p@srheight{\@p@sheight} \fi
		\if@rwidth \else \edef\@p@srwidth{\@p@swidth} \fi
}
\def\compute@sizes{
	\compute@bb
	\if@scalefirst\if@angle
	\if@width
	   \in@hundreds{\@p@swidth}{\@bbw}{\ps@bbw}
	   \edef\@p@swidth{\@result}
	\fi
	\if@height
	   \in@hundreds{\@p@sheight}{\@bbh}{\ps@bbh}
	   \edef\@p@sheight{\@result}
	\fi
	\fi\fi
	\compute@handw
	\compute@resv
	\if@scale
	   \if@verbose
	      \ps@typeout{(scaling by \@p@sscale)}%
	   \fi
	   \ps@scaleinhundreds{\@p@swidth}%
	   \ps@scaleinhundreds{\@p@sheight}%
	   \ps@scaleinhundreds{\@p@srwidth}%
	   \ps@scaleinhundreds{\@p@srheight}%
	\fi
}

\def\psfig#1{\vbox {
	\ps@init@parms
	\parse@ps@parms{#1}
	\compute@sizes
	\ifnum\@p@scost<\@psdraft{
		\special{ps::[begin] 	\@p@swidth \space \@p@sheight \space
				\@p@sbbllx \space \@p@sbblly \space
				\@p@sbburx \space \@p@sbbury \space
				startTexFig \space }
		\if@angle
			\special {ps:: \@p@sangle \space rotate \space} 
		\fi
		\if@clip{
			\if@verbose{
				\ps@typeout{(clip)}
			}\fi
			\special{ps:: doclip \space }
		}\fi
		\if@prologfile
		    \special{ps: plotfile \@prologfileval \space } \fi
		\if@decmpr{
			\if@verbose{
				\ps@typeout{psfig: including \@p@sfile.Z \space }
			}\fi
			\special{ps: plotfile "`zcat \@p@sfile.Z" \space }
		}\else{
			\if@verbose{
				\ps@typeout{psfig: including \@p@sfile \space }
			}\fi
			\special{ps: plotfile \@p@sfile \space }
		}\fi
		\if@postlogfile
		    \special{ps: plotfile \@postlogfileval \space } \fi
		\special{ps::[end] endTexFig \space }
		\vbox to \@p@srheight true sp{
			\hbox to \@p@srwidth true sp{
				\hss
			}
		\vss
		}
	}\else{
		\if@draftbox{		
			\hbox{\frame{\vbox to \@p@srheight true sp{
			\vss
			\hbox to \@p@srwidth true sp{ \hss \@p@sfile \hss }
			\vss
			}}}
		}\else{
			\vbox to \@p@srheight true sp{
			\vss
			\hbox to \@p@srwidth true sp{\hss}
			\vss
			}
		}\fi

	}\fi
}}
\psfigRestoreAt
\let\@=\LaTeXAtSign